\documentstyle[epsf,epic,eepic]{EuroPhys}

\def\And{{\rm and\ }}

\def\stars{\bigskip\centerline{***}\medskip}

\newif\ifboo \boofalse

\def\hlfplane{\frac 12\infty}

\def\be{\begin{equation}}
\def\ee{\end{equation}}

\begin{document}
\euro{xx}{x}{1-7}{2002}
\Date{today}
\shorttitle{B. BERCHE, A. FARI$\tilde{\rm N}$AS \And R. PAREDES: 
  CORRELATIONS IN THE 2D $XY$ MODEL}

\title{Correlations in the low-temperature phase of
  the two-dimensional $XY$ model}

\author{B. Berche$^{(*)}$, A.I. Fari$\tilde{\sc n}$as Sanchez$^{(**)}$ 
  \And R. Paredes V.$^{(**)}$}
\institute{$^{(*)}$ Laboratoire de Physique des Mat\'eriaux, UMR CNRS 7556,\\ 
  Universit\'e Henri Poincar\'e, Nancy
  1,\\ B.P. 239,
  F-54506  Vand\oe uvre les Nancy Cedex, France\\
  $^{(**)}$ Centro de F\'\i sica, 
  Instituto Venezolano de Investigaciones Cient\'\i ficas,\\
  Apartado 21827, Caracas 1020A, Venezuela\\}

\pacs{
  \Pacs{05}{50.+q}{Lattice theory and statistics; Ising problems}
  \Pacs{75}{10}{General theory and models of magnetic ordering}
}

\maketitle

\begin{abstract}
  Monte Carlo simulations of the two-dimensional $XY$ model are performed in
  a square geometry with fixed boundary conditions. Using a conformal mapping 
  it is very easy to deduce the
  exponent $\eta_\sigma(T)$ of the order parameter correlation
  function at any temperature in the critical  
  phase of the model.
  The temperature behaviour of $\eta_\sigma(T)$ is obtained
  numerically with a good accuracy up to the Kosterlitz-Thouless transition temperature.
  At very low temperatures, a good agreement is found with Berezinskii's
  harmonic approximation. 
  Surprisingly, we show some evidence that there are no logarithmic corrections to the behaviour of the
  order parameter density profile (with symmetry breaking surface fields) at the 
  Kosterlitz-Thouless transition temperature.
\end{abstract}
\vspace{-0.75cm}


\noindent The two-dimensional classical $XY$ model is very famous
in statistical physics, both for fundamental reasons, as describing for example classical Coulomb gas or
fluctuating surfaces, and because of its relevance for
the description of real systems, e.g. Helium super-fluid films. The model 
undergoes a standard temperature-driven paramagnetic to ferromagnetic phase transition in $d>2$, 
characterised e.g. by a power-law divergence of the correlation length near criticality,
$\xi\sim|t|^{-\nu}$,
(see  e.g. Ref.~\cite{HasenbuschToeroek99})
while it exhibits a rather different behaviour in two dimensions, 
where the existence of a low-temperature phase with conventional
long range order is precluded 
by the Mermin-Wagner theorem~\cite{MerminWagner66}. 
Indeed there is no  spontaneous magnetisation at any nonzero temperature in one or two-dimensional
isotropic systems with short range interactions and having a continuous symmetry group, for example
$O(n)$ spin models for $n\ge 2$, since order would otherwise be destroyed by spin wave
excitations~\cite{Berezinskii71}.
Such systems can however display a low-temperature phase with topological order and  
the defect-mediated transition is governed by unbinding of topological defects.
At low temperature, the system is partially ordered, apart from
the topological defects (vortices in the case of the $2d$ $XY$-model) 
which appear in pairs, in increasing
number with increasing temperature. At the transition temperature, the pairs
are broken and the system becomes completely disordered. The low-temperature regime
was investigated  by Berezinskii~\cite{Berezinskii71} in the harmonic approximation
and the mechanism of unbinding of vortices was studied by
Kosterlitz and Thouless~\cite{KosterlitzThouless73,Kosterlitz74} using approximate
renormalization group methods. For reviews,
see e.g. Refs.~\cite{KosterlitzThouless78,Nelson83,ItzyksonDrouffe89,GulasciGulasci98}. 

This very peculiar topological transition is characterised by essential singularities 
when approaching the critical point from the high temperature phase,
$ \xi\sim{\rm e}^{b_\xi t^{-\nu}}$, $\chi\sim{\rm e}^{b_\chi t^{-\nu}}$,
while in the low-temperature phase, there is a line of critical points and 
the order parameter correlation function  
decays algebraically with an exponent $\eta_\sigma(T)$ which depends on the
temperature.  
Consider a square lattice with two-components spin variables 
$\vec\phi_w=(\cos\theta_w,\sin\theta_w)$  
with continuous $O(2)$ symmetry, located at the sites 
$w$ of a lattice $\Lambda$ of linear extent $L$, 
and interacting through the usual nearest-neighbour interaction
\be
        -\frac{H}{k_BT}=K\sum_{w}\sum_\mu\vec\phi_w\cdot\vec\phi_{w+\hat\mu}
        =K\sum_{w}\sum_\mu\cos(\theta_w-\theta_{w+\hat\mu}),
        \label{Ham}
\ee
where 
$\mu$ labels the directions and $\hat\mu$ is a unit vector in the $\mu-$direction.
The low-temperature or spin-wave behaviour is obtained in the harmonic
approximation, after expanding the cosine, which should be justified at sufficiently low
temperature where existence of order, at least at short range, is assumed:
\be
-\frac{H}{k_BT}\simeq 
-\frac{H_0}{k_BT}-\frac 12K\sum_{w}\sum_\mu(\theta_w-\theta_{w+\hat\mu})^2.
\ee
Within this approximation, the two-point correlator becomes
\begin{eqnarray}
\langle\vec\phi_{w_1}\cdot\vec\phi_{w_2}\rangle&
\simeq&
Z_\beta^{-1}\prod_w\int\frac{d\theta_w}{2\pi}\cos(\theta_{w_1}-\theta_{w_2})\times e^{-\frac 12K
\sum_{w}\sum_\mu(\theta_w-\theta_{w+\hat\mu})^2}\nonumber \\
&
\simeq&
|w_1-w_2|^{-1/2\pi K},
\end{eqnarray}
hence $\eta(T)=T/2\pi$.
The low-temperature phase thus has the characteristic features 
of a critical phase with local scale invariance, {\it i.e.} conformal invariance
(invariance under rotation, translation and scale transformations,
short-range interactions, isotropic scaling).
Conformal invariance is known as a powerful
framework for the description of two-dimensional critical systems~\cite{Henkel99}.
In the case of the $XY$ model, the central
charge is $c=1$ in agreement with the continuous variation of critical exponents in the low-temperature
phase. The scaling dimensions $x_{n,m}$ of the primary operators
${\cal O}_{n,m}$ depend on a parameter $\rho$ (called compactification radius),
\be
x_{n,m}=\frac 12(n^2/\rho^2+m^2\rho^2).\label{xnm}
\ee
The spin operator is identified to the conformal operator ${\cal O}_{1,0}$ 
 and the vortex operator to ${\cal O}_{0,1}$~\cite{Henkel99}. 
This is coherent with
the picture according to which in the high-temperature phase, the vortices are unbounded 
and produce the disordering of the system, the vortex chemical potential  playing the r\^ole of a 
relevant operator. Then vortices and anti-vortices begin to bind, decreasing their relevance when the 
temperature decreases to become marginal at $T_{\rm KT}$, where the condensation of defects
occurs. There, $x_{0,1}$ must be 2
(for the RG eigenvalue $y=d-x_{0,1}$ to vanish),
implying that $\rho=2$. Hence at the KT transition, the exponent of the spin-spin correlation function decay
takes the value predicted by Kosterlitz and Thouless,
 $\eta_\sigma(T_{\rm KT})=2x_{1,0}=1/4$.
The corresponding values in the critical phase below $T_{\rm KT}$
would in principle 
easily be deduced from equation~(\ref{xnm}),
but the dependence of $\rho$ on the temperature is not known. 
Unfortunately, not much is known in the intermediate regime
between the spin wave approximation
at low temperature and the Kosterlitz-Thouless results at the topological transition
and in fact the precise determination of the critical behaviour of the two-dimensional 
$XY$ model in the critical
phase remains a challenging problem. Many results were obtained using 
Monte Carlo simulations (see e.g. 
Refs.~\cite{FernandezEtal86,BifferalePetronzio89,Wolff89,GuptaBaillie92,
Janke97,KennaIrving97})
at $T_{\rm KT}$ or high-temperature series expansions~\cite{
ButeraComi94},
but the analysis was made difficult due to the existence of logarithmic corrections, e.g.
$\chi\sim\xi^{2-\eta_\sigma}(\ln \xi)^{-2r}$ 
in the high-temperature
regime, or $\langle\vec\phi_{w_1}\cdot\vec\phi_{w_2}\rangle
\simeq|w_1-w_2|^{-\eta_\sigma}(\ln |w_1-w_2|)^{\eta_\sigma/2}$ at $T_{\rm KT}$ 
exactly~\cite{ItzyksonDrouffe89} 
and the value of $\eta_\sigma$ at $T_{\rm KT}$ was a bit controversial as shown in table~1
of reference~\cite{KennaIrving97}.
The resort to large-scale simulations was then needed in order to confirm this picture~\cite{Janke97}.


In the following, we use a rather different approach which does not require so extensive
simulations. Assuming that the low-temperature phase exhibits all the characteristics of a critical
phase with conformal symmetry, we simply use the covariance law of $n-$point correlation
functions under the mapping of a two-dimensional system confined inside a square onto the half-infinite
plane. The scaling dimensions are then obtained through a simple power-law fit where the shape
effects are encoded in the conformal mapping. This is the crucial point, since then even very small
systems are well adapted to such fits, apart from irrelevant lattice effects.
We note that mappings inside a square and various 
other geometries  have been considered already, e.g.
the moments of the magnetization~\cite{BurkhardtDerrida85} and structure factors~\cite{KlebanEtal86} 
in the Ising model have been calculated in square systems.

The order parameter correlation function in a square system,
$\langle\vec\phi_{w_1}\cdot\vec\phi_{w_2}\rangle$,
should in principle lead to  
the determination of critical exponents, but practically, it
is not of great help 
since strong surface effects occur which modify the large-distance power-law 
behaviour.  The correlation function is supposed to obey a scaling
form which reproduces the expected power-law behaviour in the thermodynamic 
limit, for example 
$ \langle\vec\phi_{w_1}\cdot\vec\phi_{w_2}\rangle =|w_1-w_2|^{-\eta_\sigma}
  f_{\rm sq.}(w_1/L,w_2/L)$,
where $f_{\rm sq.}$ encodes shape effects in a very
complicated manner.
Conformal invariance provides an efficient technique to avoid these shape
effects, or at least, enable to include explicitly the shape
dependence in the functional expression of the correlators through the conformal
covariance transformation under a mapping $w(z)$:  
\be
\langle\vec\phi_{w_1}\cdot\vec\phi_{w_2}\rangle=|w'(z_1)|^{-x_\sigma}|w'(z_2)|^{-x_\sigma}
\langle\vec\phi_{z_1}\cdot\vec\phi_{z_2}\rangle
\label{covconf}
\ee
with $\eta_\sigma=2x_\sigma$.
In the semi-infinite geometry $z=x+iy$ (the free surface being defined by the $x$ axis),  
the two-point correlator is fixed up
to an unknown scaling function (apart from some asymptotic limits
implied by scaling). Fixing one point $z_1$ close to the free surface 
($z_1=i$) of
the half-infinite plane, and leaving the second point $z_2$ explore the
rest of the geometry, the following behaviour is expected:
$\langle \vec\phi_{z_1}\cdot\vec\phi_{z_2}\rangle_{\hlfplane}
\sim (y_1-y_2)^{-x_\sigma}\psi(\omega),$
where the dependence on $\omega={y_1y_2}/{\mid z_1-z_2\mid^2}$ of the 
universal scaling function $\psi$ is constrained
by the special conformal transformation~\cite{Cardy84}.
In order to get a functional expression of the correlation function
inside the square geometry $w$, one simply has to use the mapping $w(z)$ which
realizes the conformal transformation of the half-plane $z=x+iy$ ($0\le y<\infty$) 
inside a square $w=u+iv$ of size
$L\times L$ ($-L/2\le u\le L/2$, $0\le v\le L$)
with free boundary conditions
along the four edges. This is realized by a Schwarz-Christoffel transformation~\cite{LavrentievChabat}
\be
w(z)={L\over 2{\rm K}}{\rm F}(z,k),
\quad z={\rm sn}\left({2{\rm K}w\over L}\right).
\label{eq-SchChr}
\ee
Here, $F(z,k)$ is the elliptic integral of the 
first kind, ${\rm sn}\ \! (2{\rm K}w/ L)$ 
the Jacobian elliptic sine, ${\rm K}=K(k)$ the
complete elliptic integral of the first kind, and the modulus $k$ depends on the aspect ratio
of $\Lambda$  and is here solution
of $K(k)/K(\sqrt{1-k^2})=\frac 12$.
Using the mapping~(\ref{eq-SchChr}), the local rescaling factor in 
equation~(\ref{covconf}) is obtained,
$w'(z)=\frac{L}{2{\rm K}}[(1-z^2)(1-k^2z^2)]^{-1/2}$,
and inside the square, keeping $w_1$ fixed, the two-point correlation function 
becomes (see e.g. Ref.~\cite{ChatelainBerche99})
\be
\langle\vec\phi_{w_1}\cdot\vec\phi_w\rangle_{\rm sq.} \sim {\underbrace{
\left(\Im [z].
/\sqrt{
|1-z^2|.|1-k^2z^2|
}
\right)
}_{\kappa(w)}}^{-x_\sigma}\psi(\omega),
\label{eqkappa}\ee
with $z(w)$ given by equation~(\ref{eq-SchChr}). 
This expression is correct up to a constant amplitude determined by
$\kappa(w_1)$ which is kept fixed, but the function $\psi(\omega)$ is still
varying with the location of the second point, $w$. 
In order to cancel the r\^ole of the unknown scaling function,
it is more convenient to work with a density profile $m(w)$ in the presence
of symmetry breaking surface fields $\vec h_{\partial\Lambda}$ 
on the boundary $\partial\Lambda$ of the lattice 
$\Lambda$. This 
one-point correlator 
scales
in the half-infinite geometry 
$m(z)=\langle\vec\phi_z\cdot\vec h_{\partial\Lambda(z)}\rangle_{\hlfplane} 
= {\rm const\times} y^{-x_\sigma}$
and it maps onto
\be
m(w)=\langle\vec\phi_w\cdot\vec h_{\partial\Lambda(w)}\rangle_{\rm sq.} =
        {\rm const\times}[\kappa(w)]^{-x_\sigma}
\label{eqParamOrder}        \ee
where the function $\kappa(w)$ defined in equation~(\ref{eqkappa}) again comes from the mapping.


The application of the simple power-law in equation~(\ref{eqParamOrder}) requires a relatively
precise numerical determination of the order parameter profile of the $2d$ $XY$ model
confined inside a square with fixed boundary conditions playing the r\^ole of
ordering surface fields $\vec h_{\partial\Lambda(w)}$. In practice, no magnetic field is applied,
but the symmetry is broken by keeping the  boundary spins of the square fixed, 
e.g. $\vec\phi_w=(1,0), \forall w\in\partial\Lambda$ during the Monte Carlo 
simulations (in principle no update of these spins is performed).
In the low-temperature phase, local Metropolis updates of single spins are known  
to suffer from the critical slowing down, since the system contains arbitrarily large clusters
in which the spin orientations are strongly correlated. Statistically independent
configurations can be obtained by local iteration rules only after a long
dynamical evolution which needs a huge number of MC steps, sometimes beyond nowadays
computers capability.  The resort
to cluster update algorithms (like Wolff algorithm) is more convenient.
The central idea of cluster algorithms is the identification of 
clusters of sites using a bond percolation process connected to the spin
configuration. The spins of the clusters are then independently updated.
A cluster algorithm is particularly efficient if the percolation 
threshold coincides with the transition point of the spin model, 
which guarantees that clusters of all sizes will be updated in a single
MC sweep. The percolation process associated to Ising or Potts models is the random
graph model of the Fortuin-Kasteleyn representation and its threshold is known to coincide
with the critical point of the spin model. For the $XY$ model, the bonds are
introduced in the Wolff algorithm through the definition of Ising variables defined by the sign
of the projection of the spin variables along some random direction. The percolation threshold
for these bonds coincides with the Kosterlitz-Thouless point~\cite{DukovskiMachtaChayes01}, 
which guarantees the efficiency of the
Wolff cluster updating scheme~\cite{Wolff89} at $T_{\rm KT}$. 
In the low-temperature phase, this algorithm
could be less efficient, but nevertheless preferable to a local updating.
When one uses particular boundary
conditions, with fixed spins along some 
surface as required here, the Wolff algorithm should
become less efficient, since close to criticality the unique cluster will often
reach the boundary and no update is made in this case. To prevent this, we use the symmetry
properties of the Hamiltonian~(\ref{Ham}). Even when the cluster reaches the fixed boundaries
$\partial\Lambda(w)$, 
it is updated, and the order parameter profile is then measured with respect to the common
new direction  of the boundary
spins, $m(w)=\langle\vec\phi_w\cdot\vec \phi_{\partial\Lambda}\rangle_{\rm sq.}$.
The new configuration reached would thus correspond - after a global rotation of all the spins 
of the system to re-align
the boundary spins in the $(1,0)$ direction - to a
a new configuration of equal total energy
and thus the same statistical weight as the one actually produced.
This technique  eventually leads to a new update at each iteration: when the cluster does not touch the 
boundaries, all the spins belonging to the cluster are updated, and in the other case the situation becomes
equivalent 
to an update of
all the spins but those inside the cluster. At a percolation threshold, interior and exterior of a percolating
cluster are both large and intricated and the algorithm has short time correlations.
After averaging over the `production sweeps', one gets a characteristic smooth profile.

\begin{figure} [h]
        \epsfxsize=11.5cm
        \begin{center}
        \mbox{\epsfbox{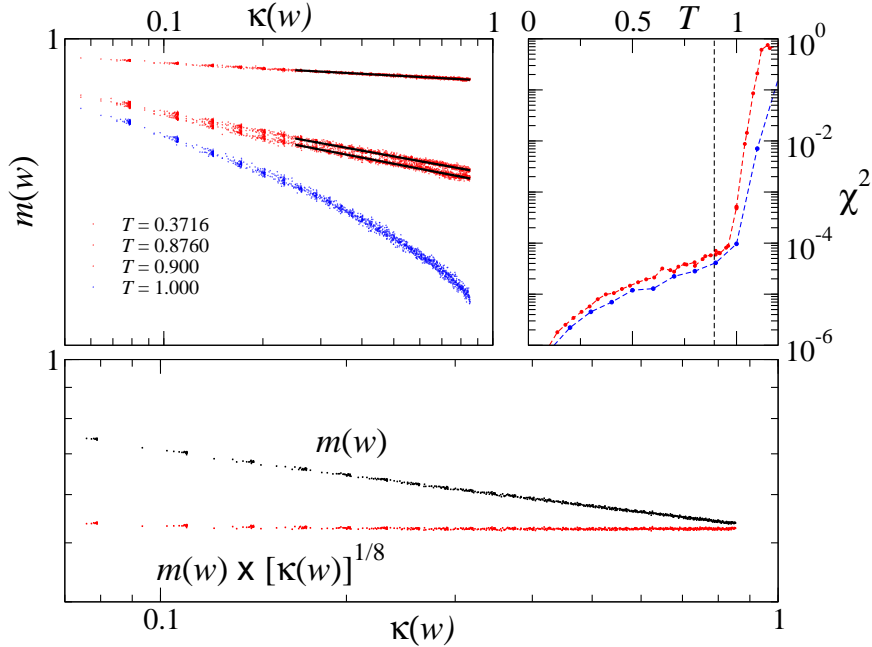}\qquad}
        \end{center}
        \caption{Monte Carlo simulations of the $2d$ $XY$ model inside a square
        $100\times 100$ spins ($10^6$ MCS/spin after cancellation of $10^6$ for
        thermalization, cluster update algorithm). Top: on the left, the 
        figure shows the rescaled order parameter density against the rescaled
        distance on a log-log scale at  different temperatures below the Kosterlitz-Thouless
        transition temperature and slightly above. On the right, we show a plot of the $\chi^2$ per d.o.f. 
        of the power-law fits as 
          a function of the temperature for $L=48$ and $100$. 
          The sharp change of behaviour is a signature of the location
        of the Kosterlitz-Thouless transition, above which the functional expression 
        (\ref{eqParamOrder}) of the density profile
      is no longer correct.
      Bottom: plot of the density profile $m(w)$ with symmetry breaking surface field and of the rescaled
          quantity $m(w)\times[\kappa(w)]^{1/8}$ {\it vs} $\kappa(w)$ on a log-log scale (exactly 
          at the Kosterlitz-Thouless transition temperature). The rescaled profile remains constant with
        a very good accuracy, indicating the absence of logarithmic correction for this quantity.}
        \label{fig:1}  \vskip -0cm
\end{figure}
\begin{figure} [h]
\vspace{0.2cm}
        \epsfxsize=11cm
        \begin{center}
        \mbox{\epsfbox{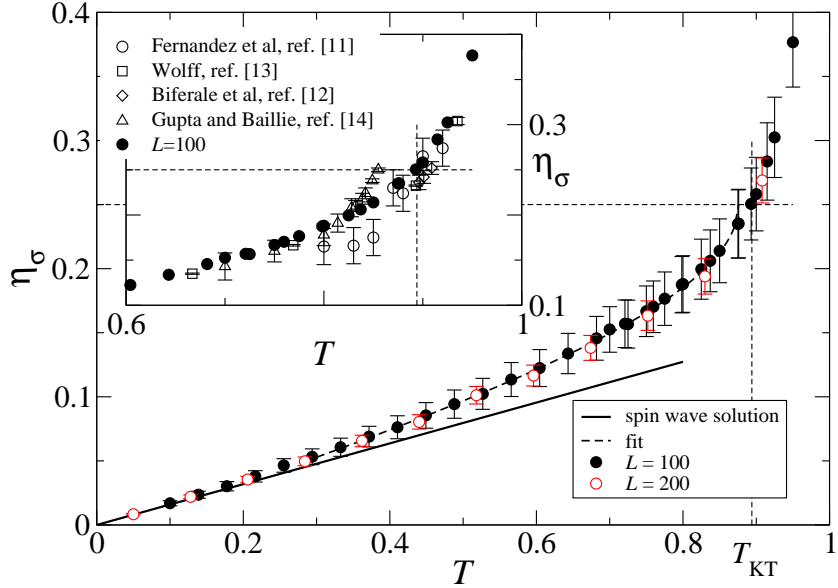}\qquad}
        \end{center}
        \caption{Correlation decay exponent of the $2d$ $XY$ 
          model as a function of the temperature. The insert
        shows a comparison with other results (in the vicinity of the KT point) found in the literature.}
        \label{fig:2}  \vskip -0cm
\end{figure}

A log-log plot of $m(w)$ with respect to the reduced variable $\kappa(w)$ is shown in figure~\ref{fig:1}
at two different temperatures below $T_{\rm KT}\simeq 0.893(1)$ (the value of $T_{\rm KT}$ is taken
according to reference~\cite{GuptaBaillie92}), 
roughly at $T_{\rm KT}$, and one temperature
slightly above. 
The first message of the plot is the confirmation of the functional form of
equation~(\ref{eqParamOrder}). One indeed observes a very good data collapse of the $L^2$ points
onto a single power-law  master curve (a straight line on this scale). 
The next information is the rough confirmation of the value
of the critical temperature, since above $T_{\rm KT}$, the master curve is no longer a straight line,
indicating that the corresponding decay in the half-infinite geometry differs from 
a power-law as it should in the high-temperature phase. One should
nevertheless mention that this is not a technique adapted to a precise determination of $T_{\rm KT}$,
since at $T=0.9$ for example, the master curve is hardly distinguishable from a straight line. 
One can for example compute the $\chi^2$ per d.o.f. as a function of temperature. It has a very
small value, indicating the high quality of the fit, in the low-temperature phase and then increases 
significantly above $T_{\rm KT}$ when the behaviour of the density profile is no longer algebraic.
The change of behaviour in the curve gives an approximate location of the Kosterlitz-Thouless
transition temperature, and the larger the system size, the better this estimate. 
This is illustrated in figure~\ref{fig:1}.

It is tempting to study the r\^ole of logarithmic corrections exactly at the Kosterlitz-Thouless point.
For that purpose, we produce data at a temperature $T=T_{\rm KT}$ and, assuming that 
logarithmic corrections should affect the density profile with symmetry breaking surface field
in the half-infinite geometry, we make a plot of $m(w)\times[\kappa(w)]^{1/8}$ {\it vs} $\kappa(w)$.
Surprisingly, this leads very accurately to a constant, as shown in figure~\ref{fig:1}.
This is a clear evidence that the order parameter profile with fixed boundary conditions
displays a pure algebraic decay at the Kosterlitz-Thouless transition point. The decay exponent,
if not fixed, but determined numerically, leads to a quite good result 
(for example for a size $L=100$, we get 
$\eta_\sigma(T_{\rm KT})=0.250(28)$).

It is thus possible to obtain the scaling
dimension $x_\sigma(T)$ or the two-point correlation function exponent 
$\eta_\sigma(T)=2x_\sigma(T)$, in the whole critical region $T\le T_{\rm KT}$
as shown in figure~\ref{fig:2}.
The most interesting result obtained here 
is perhaps the fact that the values of  the exponent $\eta$ are 
very easy to obtain, in spite of the reputed difficulty of this problem from MC studies. 
For example the results obtained with a system
of size as small as $48\times 48$ are already very satisfying,  and to
get 10 values of $T$ ($10^5$  MCS/spin for thermalization $+10^6$ MCS/spin for computation) 
corresponds to a quite reasonable amount of 11 hours of CPU time
on a standard PC with 733 MHz processor (the algorithm is more efficient at
$T_{\rm KT}$ than below) . For the largest size considered here, $L=200$, a simulation
at the Kosterlitz-Thouless temperature
needs 18 hours ($10^6+10^6$ MCS/spin) which is also very fast compared to today's standard extensive
simulations where the natural time unit is the CPU-year.
We mention here that we do not reach the accuracy of the very precise 
numerical determination of the magnetic scaling dimension
(using transfer matrix techniques) by Bl\"ote and 
Nienhuis~\cite{BloteNienhuis89}.

Once the values of these scaling dimensions are known, we can deduce numerically the
temperature-behaviour of the compactification parameter $\rho$, and thus perhaps
conjecture the numerical values of other scaling dimensions associated to different operators.
Up to the authors knowledge, such an approach has not yet been done and it could be
promising for a better understanding of the critical phase of the Kosterlitz-Thouless
transitions. 
From the numerical data, $\eta_\sigma(T)$ is easily expanded close to the 
KT point $T\to T_{\rm KT}^-$, showing a leading square root behaviour~\cite{Kosterlitz74},
\be
\eta_\sigma\sim \frac 14 -0.231(T_{\rm KT}-T)^{1/2}-0.050 (T_{\rm KT}-T)^{3/2},
\ee
shown as a fit in figure~\ref{fig:2}
and from which it is easy to write the first terms of an expansion of the compactification radius, 
$\rho\simeq 2[1+0.462 (T_{\rm KT}-T)^{1/2}+0.320 (T_{\rm KT}-T)+\dots]$ 
when $T\to T_{\rm KT}^-$.
This leads in principle to an approximate value of the other scaling dimensions through 
equation~\ref{xnm}.

\stars
BB would like to thank Christophe Chatelain for  stimulating discussions and  Dragi
Karevski for a critical reading of the manuscript.



\vskip-12pt
\begin{thebibliography}{99}
\def\paper#1#2#3#4#5{{\sc #1}, {\it #2}\ {\bf #3}, #4 (#5).}

\bibitem{HasenbuschToeroek99}\paper{M. Hasenbusch \And T. T\"or\"ok}{J. Phys. A}{32}{6361}{1999}
\bibitem{MerminWagner66} \paper{N.D. Mermin \And H. Wagner}{Phys. Rev. Lett.}{22}{1133}{1966}
\bibitem{Berezinskii71}\paper{V.L. Berezinskii}{Sov. Phys. JETP}{32}{493}{1971}
\bibitem{KosterlitzThouless73}\paper{J.M. Kosterlitz \And D.J. Thouless}
        {J. Phys. C}{6}{1181}{1973}
\bibitem{Kosterlitz74}\paper{J.M. Kosterlitz}{J. Phys. C}{7}{1046}{1974}
\bibitem{KosterlitzThouless78}\paper{J.M. Kosterlitz \And D.J. Thouless}
        {Prog. Low Temp. Phys}{78}{371}{1978}
\bibitem{Nelson83} {\sc D.R. Nelson}, in {Phase Transitions and Critical Phenomena}, ed. by
  {\sc C. Domb \And J.L. Lebowitz}, Academic Press, London 1983, p.~1.
\bibitem{ItzyksonDrouffe89} {\sc C. Itzykson \And J.M. Drouffe}, {Statistical field theory},
  Cambridge University Press, Cambridge 1989, vol. 1.
\bibitem{GulasciGulasci98}\paper{Z. Gul\'acsi \And M. Gul\'acsi}{Adv. Phys.}{47}{1}{1998}
\bibitem{Henkel99} {\sc M. Henkel}, {Conformal Invariance and Critical Phenomena}, 
  Springer, Heidelberg 1999.
\bibitem{FernandezEtal86}\paper{J.F. Fern\'andez, M.F. Ferreira \And J. Stankiewicz}
  {Phys. Rev. B}{34}{292}{1986}
\bibitem{BifferalePetronzio89}\paper{L. Bifferale \And R. Petronzio}{Nucl. Phys. B}{328}{677}{1989}
\bibitem{Wolff89}\paper{U. Wolff}{Nucl. Phys. B}{322}{759}{1989}
\bibitem{GuptaBaillie92} \paper{R. Gupta \And C.F. Baillie}{Phys. Rev. B}{45}{2883}{1992}
\bibitem{Janke97}\paper{W. Janke}{Phys. Rev. B}{55}{3580}{1997}
\bibitem{KennaIrving97}\paper{R. Kenna \And A.C. Irving}{Nucl. Phys. B}{485 [FS]}{583}{1997}
\bibitem{ButeraComi94}\paper{P. Butera \And M. Comi}{Phys. Rev. B}{50}{3052}
  {1994}
\bibitem{BurkhardtDerrida85} \paper{T.W. Burkhardt  \And  B. Derrida}{Phys. Rev. B}{32}{7273}{1985}
\bibitem{KlebanEtal86} \paper{P. Kleban, G. Akinci, R. Hemtschke  \And  K.R. 
   Brownstein}{J. Phys. A}{19}{437}{1986}
\bibitem{Cardy84}  \paper{J. L. Cardy} {Nucl. Phys. B} {240 [FS12]} {514}{1984} 
\bibitem{LavrentievChabat} {\sc M. Lavrentiev \And B. Chabat}, {M\'ethodes
   de la th\'eorie des fonctions d'une variable complexe}, Mir, Moscou 1972, 
   Chap. VII.
\bibitem{ChatelainBerche99} \paper{C. Chatelain \And B. Berche}{Phys. Rev. E}{60}
  {3853}{1999}
\bibitem{DukovskiMachtaChayes01} \paper{I. Dukovski, J. Machta \And L.V. Chayes}
        {Phys. Rev. E}{65}{026702}{2002}
\bibitem{BloteNienhuis89}\paper{H.W.J. Bl\"ote \And B. Nienhuis}{J. Phys. A}{22}{1415}{1989}
\end{thebibliography}
\end{document}